\documentclass[sigconf]{acmart}
\usepackage{graphicx} %

\title{Towards Shift-Up: A Framework and a Prestudy on High-Value Activities in GenAI Native Software Development}

\author{Vlad Stirbu}
\affiliation{%
  \institution{University of Jyväskylä}
  \city{Jyväskylä}
  \country{Finland}}
\email{vlad.a.stirbu@jyu.fi}

\author{Mateen Ahmed Abbasi}
\affiliation{%
  \institution{University of Jyväskylä}
  \city{Jyväskylä}
  \country{Finland}}
\email{mateen.a.abbasi@jyu.fi}

\author{Teerath Das}
\affiliation{%
  \institution{University of Jyväskylä}
  \city{Jyväskylä}
  \country{Finland}}
\email{teerath.t.das@jyu.fi}

\author{Jesse Haimi}
\affiliation{%
  \institution{University of Jyväskylä}
  \city{Jyväskylä}
  \country{Finland}}
\email{jesse.k.haimi@jyu.fi}

\author{Niko Iljin}
\affiliation{%
  \institution{University of Jyväskylä}
  \city{Jyväskylä}
  \country{Finland}}
\email{niko.s.iljin@jyu.fi}

\author{Pyry Kotilainen}
\affiliation{%
  \institution{University of Jyväskylä}
  \city{Jyväskylä}
  \country{Finland}}
\email{pyry.kotilainen@jyu.fi}

\author{Petrus Lipsanen}
\affiliation{%
  \institution{University of Jyväskylä}
  \city{Jyväskylä}
  \country{Finland}}
\email{petrus.i.lipsanen@jyu.fi}

\author{Niko Mäkitalo}
\affiliation{%
  \institution{University of Jyväskylä}
  \city{Jyväskylä}
  \country{Finland}}
\email{niko.k.makitalo@jyu.fi}

\author{Maiju Sipilä}
\affiliation{%
  \institution{University of Jyväskylä}
  \city{Jyväskylä}
  \country{Finland}}
\email{maiju.a.sipila@student.jyu.fi}

\author{Venla Veijalainen}
\affiliation{%
  \institution{University of Jyväskylä}
  \city{Jyväskylä}
  \country{Finland}}
\email{venla.m.veijalainen@jyu.fi}

\author{Tommi Mikkonen}
\affiliation{%
  \institution{University of Jyväskylä}
  \city{Jyväskylä}
  \country{Finland}}
\email{tommi.j.mikkonen@jyu.fi}

\date{September 2025}

\begin{document}

\acmConference[]{}{} {}
\acmYear{2026}
\copyrightyear{2026}

\begin{abstract}
Generative AI (GenAI) has significantly influenced software engineering. Associated tools have created a shift in software engineering, where specialized agents, based on user-provided prompts, are replacing human developers. In this paper, we propose a framework for GenAI native development that we call \textit{shift-up}, which helps software teams focus on high-value work while being supported by GenAI. Furthermore, we also present a preliminary study testing these ideas with current GenAI tools. Towards the end of the paper, we propose future research goals to study shift-up in more detail.
\end{abstract}


\maketitle

\section{Introduction}

The emergence of Generative AI (GenAI) has had a profound impact on software engineering \cite{belzner2023large}.  Such models are becoming more and more capable of aiding different software engineering tasks, whether it be requirements engineering, code generation or documentation \cite{nguyen2024generative}.
This has led to their widespread adoption in software development processes \cite{stray2025generative}.

The adoption of GenAI tools in software development activities has created a shift in which actual software development is no longer performed by humans but by GenAI, based on user-provided prompts. As the GenAI tools become more capable, we expect that they will be able to address many other software development activities, too.

In this paper we propose a framework that facilitates the development of native GenAI technologies that enable the software development teams to focus on high-value activities while being assisted by a multitude of expert agents that assist them. We call this \textit{shift-up} to highlight the shift in the level of abstraction that the developers work at. In addition, we present a prestudy where the ideas of the framework are probed with today's GenAI tools.

\section{Background and motivation}
\label{sec:background}

\subsection{Software development methodologies}

The purpose of the \textbf{design control} process \cite{fda-design-control} 
is to promote a well-designed development development process that includes traceability between the input and output of the process at different stages of the process. Starting from \textit{user needs} that convert into \textit{ design input}, continuing with the \textit{design process} transforms the inputs into \textit{ design output}, and finally forms the resulting \textit{ medical device}. In addition, the reviews during each step of the process verify and validate that the requirements are met by the implementation, reducing the possibility of design and implementation defects. Design control provides a framework for governing and tracing the design process so that what is built meets the requirements, is testable, and can be maintained reliably.

\textbf{V-Model} \cite{vmodel} is a structured development approach that aligns design phases with corresponding verification and validation activities. Requirements are refined through analysis and architecture into preliminary and detailed designs, which guide coding, prototyping, and engineering. Each design phase is paired with tests at the unit, integration, and acceptance levels, ensuring systematic verification and validation before release. This alignment operationalizes design control, making the V-Model particularly suited for domains demanding compliance, safety, and reliability.

\begin{figure}
    \centering
    \includegraphics[width=0.95\linewidth]{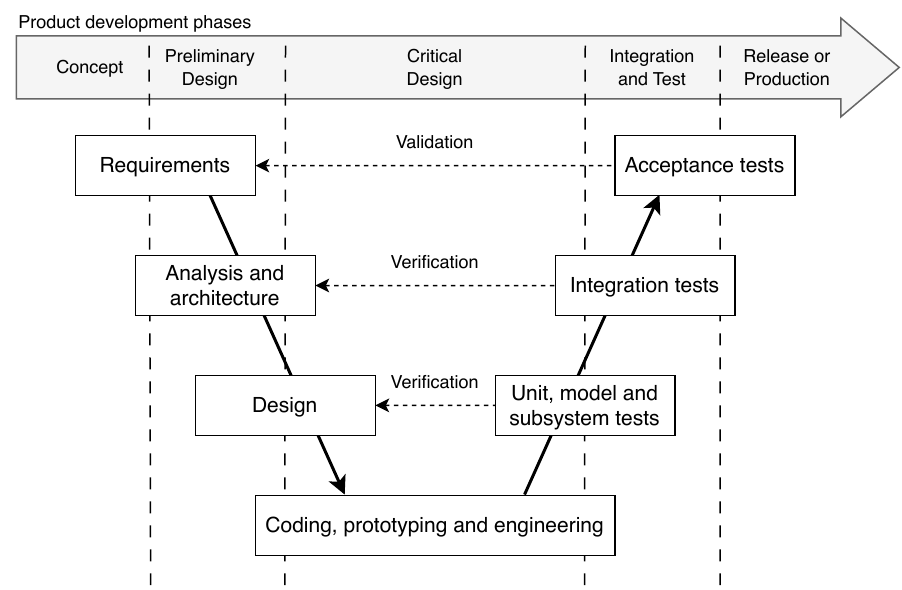}
    \caption{V-Model of development aligned with the product development phases (adapted from \cite{vmodel})}
    \label{fig:vmodel}
\end{figure}

\subsection{Software engineering best practices}

\textbf{Domain-Driven Design} (DDD) \cite{evans2003ddd} is a software development methodology that places the domain model at the heart of system design. It fosters close collaboration between domain experts and developers, using a \textit{ubiquitous language} to ensure requirements, models, and code reflect a shared understanding of the problem space. Systems are organized into \textit{bounded contexts}, each representing a consistent subdomain with its own model, which reduces coupling and clarifies responsibilities. %
By aligning architecture and implementation with evolving domain knowledge, DDD enables flexible adaptation, clearer stakeholder communication, and sustainable long-term system evolution.

\textbf{C4 model} \cite{c4} is a hierarchical method for describing software architecture that enables stakeholders to view systems at four levels of abstraction, supporting both high-level overviews and implementation‐level detail. The \textit{context} level shows the system under consideration in its environment, the \textit{container} level decomposes it into applications or services, the \textit{component} level breaks down those containers into logically cohesive modules, and the \textit{code} level optionally focuses on class or function structures in critical parts. Developed as a lightweight and developer-friendly architectural modeling approach, the C4 model emphasizes clarity and effective communication between technical and non-technical stakeholders.

\textbf{Shift-left} \cite{shift-left} is a software development and testing principle that advocates moving activities such as testing, quality assurance, or security earlier in the development lifecycle. Traditionally, these activities were concentrated toward the end of the process, often leading to late discovery of defects, costly rework, and project delays. By shifting left, teams integrate practices like unit testing, continuous integration, static code analysis, and automated security checks directly into the development workflow. This proactive approach enables earlier detection of errors, fosters rapid feedback loops, and reduces the overall cost of fixing issues. The activities are performed by multidisciplinary teams that have expertise in both development and operations (DevOps) \cite{debois2011devops}. Complementary to this is the \textbf{shift-right} approach \cite{shift-right}, which extends validation into deployment and operational phases, using practices such as canary releases, A/B testing, chaos engineering, and continuous monitoring. Together, shift-left and shift-right ensure that quality, reliability, or compliance are not only built into the product from the outset, but also continuously validated under real-world conditions.

\textbf{Architecture Decision Records} (ADRs) \cite{nygard2011adr} are lightweight documents that capture key architectural choices, including the context of the decision, the alternatives considered, the rationale behind the selected option, and the consequences of adopting it, both positive and negative. They provide a structured, traceable history of architectural evolution that supports transparency, knowledge sharing, and maintainability, especially in collaborative or distributed teams. Unlike upfront design documents, ADRs evolve incrementally with the system, making them well suited for agile and DevOps practices where architecture emerges iteratively \cite{zimmermann2015adr}.

\textbf{Executable requirements} represent a paradigm in which requirements are expressed in a form that is both human and machine-readable, enabling direct validation against the implementation \cite{fuchs99attemptoControlled}. Unlike traditional textual specifications, which are prone to ambiguity and misinterpretation, executable requirements are encoded as models, formal specifications, or testable scenarios that can be executed to verify compliance throughout the development lifecycle. Approaches such as behavior-driven development (BDD), where natural language requirements are linked to automated acceptance tests \cite{solis2011bdd}. Furthermore, model-based specification formalize requirements in state machines or temporal logic \cite{jackson2012alloy}. By embedding verification directly into the requirements themselves, executable requirements foster continuous alignment between stakeholder expectations and the software behavior, reduce defects, and serve as living documentation that evolves alongside the software. This approach has gained particular traction in safety-critical and agile contexts, where traceability and early validation are essential.

\subsection{GenAI: from chat-bots to AI workflows}

\textbf{GenAI} tools represent a class of machine learning systems that can generate text, code, images, or other content by learning patterns from large volumes of training data. The most common examples are large language models (LLMs) like GPT4~\cite{openai2024gpt4technicalreport} or LLaMA~\cite{touvron2023llama}, which are trained on vast corpora of natural language to perform tasks such as summarization, translation, or dialogue. However, the paradigm is not limited to natural language: similar architectures can be trained on other modalities or domain-specific data. For instance, models trained on source code repositories can generate or refactor software, while models exposed to financial transaction data \cite{stripe-gpt}, can learn to detect fraud, optimize payments, or support automated decision-making. This versatility demonstrates that GenAI is a general framework for pattern learning and generation, where the type of knowledge captured depends on the nature of the training data.

The \textbf{Model–Context Protocol}\footnote{\href{https://modelcontextprotocol.io/specification/2025-06-18}{https://modelcontextprotocol.io/specification/2025-06-18}} (MCP) is an open standard designed to improve interoperability between AI models, development tools, and runtime environments by establishing a common protocol for exchanging structured information. Instead of treating models as isolated black boxes, MCP introduces the concept of a shared context, where models, Integrated Development Environments (IDEs), agents, or other services can communicate domain-specific knowledge in a consistent, machine-readable form. This enables developers to integrate diverse tools and models, reducing duplication of configuration and lowering the risk of misalignment between design intent and implementation. By standardizing the way context is shared, MCP supports use cases such as interactive development, documentation generation, and tool orchestration, resulting in more reliable and transparent AI-assisted workflows.

\section{Towards GenAI Native Software Development}
\label{sec:native-ai}

\subsection{Promise of GenAI and structured knowledge}

\begin{figure}
    \centering
    \includegraphics[width=0.95\linewidth]{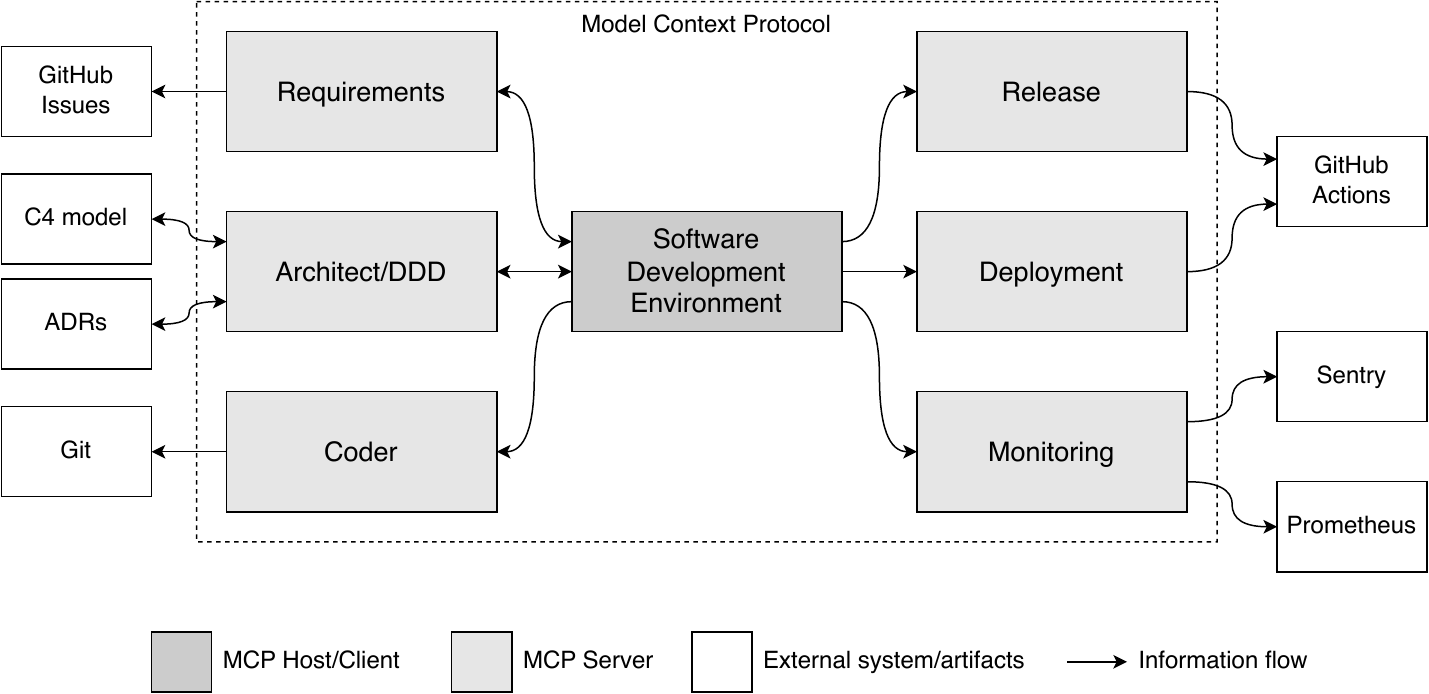}
    \caption{Native AI software development tool landscape: development environment, supporting AI agents and external systems and tools}
    \label{fig:mcp}
\end{figure}

The rise of AI-driven software development is reshaping how developers and engineers approach their work. The primary focus is shifting from writing code to defining designs and requirements, which are becoming the central development artifacts. Emerging technologies like MCP enable a new generation of tools that combine the generative power of LLMs with established engineering practices such as requirements management (e.g., GitHub Issues), architecture documentation (e.g., C4 model), and architectural evolution (e.g., ADRs). When represented as knowledge graphs \cite{LLM+KG}, these artifacts provide a structured, factual, and traceable context that complements the LLMs. This synergy enables knowledge graphs to act as domain experts, while LLMs supply the natural language fluency that makes this expertise accessible and actionable. The combination reduces hallucinations, grounds development activities within the application context, and does so without the need for additional model training. However, while the approach shows promise, there is no proven systematic way to integrate GenAI into software engineering \cite{russo2024navigating,mikkonen2025software}. Moreover, it has been argued that this process must be human-centric, and not fully automated, despite the improvements in tooling \cite{russo2024generative}. The concept is depicted in Figure \ref{fig:mcp}.

\subsection{Shift-Up}

Assisted by GenAI tools, the development team can now produce code at multiple levels of complexity, ranging from short snippets that solve a specific problem to complete modules that can be integrated and deployed into larger systems, and monitored through industry grade solutions like Sentry\footnote{\href{https://github.com/getsentry/sentry-mcp}{https://github.com/getsentry/sentry-mcp}} or Prometheus\footnote{\href{https://github.com/pab1it0/prometheus-mcp-server}{https://github.com/pab1it0/prometheus-mcp-server}}.

In the context of GenAI native software development, the shift-left term gets a renewed meaning to move activities earlier in the lifecycle by replacing code-centric practices with higher-level artifacts such as prompts, specification refinement, and automated code generation, making the developer’s intention the primary focus. Complementing this, shift-right refocuses from testing and quality assurance to deployment and operations, where GenAI systems support continuous monitoring, real-world validation, and feedback-driven evolution, ensuring robustness under production conditions.

\begin{figure}
    \centering
    \includegraphics[width=0.95\linewidth]{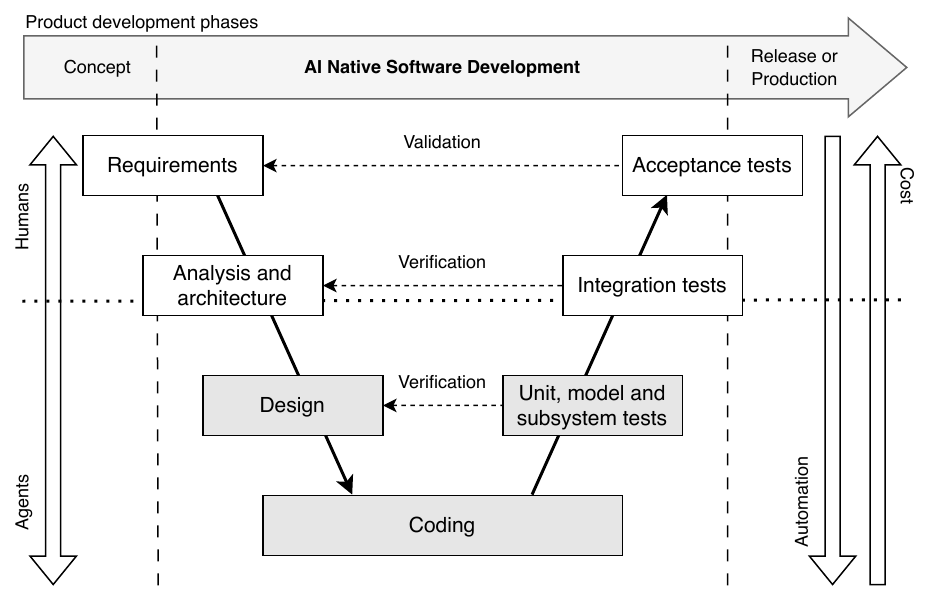}
    \caption{Shift-Up with AI native software development}
    \label{fig:ai-native}
\end{figure}

By combining these perspectives, we propose the concept of \textit{shift-up}, which fits the layered structure of the V-Model. In this GenAI native SDLC, the intermediate stages—such as detailed design, implementation, and low-level verification—are delegated to GenAI tools, allowing the human developers focus on the higher layers of the V-Model. At the top, they concentrate on requirements, architectural design, and system-level considerations, and at the bottom, they engage in acceptance testing, deployment oversight, and operational feedback. The objective of shift-up is thus to free developers from low-level implementation details and empower them to focus on strategic, creative, and domain-specific aspects of software development, while ensuring that quality and reliability are preserved throughout the lifecycle, as depicted in Figure \ref{fig:ai-native}.

Upon performing Shift-up development, humans and GenAI co-operate to solve problems in ways that are best fit to the task at hand, following a shared responsibility working model \cite{se-bots}.
The characteristic tasks of humans and GenAI are listed in Table \ref{table:characteristics}. 

\begin{table*}[t]
    \centering
    \caption{Characteristics of Shift-Up development, and the responsibilities share between the humans and the Gen AI tools}
    \label{table:characteristics}
    \begin{tabular}{lll}
    \toprule
    Development phase& Human & GenAI tools\\
    \midrule
    Requirements 
    & Requirements definition, quality control  
    & Consistency validation, refinement \\
    Analysis and architecture 
    & Quality validation, refinement
    & Architecture generation \\
    Design
    & Functional validation, refinement 
    & Design generation  \\
    Coding
    & -
    & Code generation \\
    Unit, model and subsystem tests 
    & Functional validation, refinement
    &  Test generation \\
    Integration tests 
    & Quality validation, refinement
    & Test generation, functional validation, consistency \\
    Acceptance tests
    & Quality control, monitoring
    & Test generation, Functional validation\\
    \bottomrule
    \end{tabular}
\end{table*}

\section{Prestudy: Exploring Shift-Up in Practice} \label{sec:research}

\subsection{Methodology and Research Questions}

To evaluate the Shift-Up development process, we have executed a small prestudy following the Design Science Research (DSR) \cite{dsr} methodology, in which GenAI tools has been acting as the development agents, and human actors have supervised their actions. The exact research question of the study is as follows:

\medskip
\noindent
    \textbf{RQ:} What human supervision actions make sense in GenAI native development?

\medskip
\noindent

To address the research question, a team of five undergraduate interns was assembled and tasked with developing a software system that allows users to record, comment on, and share their travel history on a map. Throughout the project, each step of the development process was supported by GenAI. Architecturally, the system follows a traditional web application model and was implemented using technologies such as React, Node, and PostgreSQL recommended by the LLMs.

The team reviewed existing literature surveys, benchmarks and leaderboards of GenAI models. Based on this review, four models were selected: Claude\footnote{https://www.anthropic.com/claude -- Claude Sonnet 4}, GPT\footnote{https://openai.com/ -- o4-mini}, DeepSeek\footnote{https://www.deepseek.com/en -- DeepSeek R1 0528}, and Gemini\footnote{https://gemini.google.com/ -- Gemini 2.5 Pro}. All four models were applied in parallel throughout the case study. Initially, each model was used independently for requirements engineering, producing a requirements document. Then, the best document was chosen and used as input for the system design phase, where the models were again employed to generate designs. The same process was followed for implementation, with models generating code based on the design documents. While the LLMs carried the main responsibility for producing requirements, design artifacts, and code, human oversight was maintained at each stage. The total development effort amounted to 10 person-months, with the team following CI/CD practices, albeit in a more flexible manner than is typical in industrial contexts.

\subsection{Results}
\label{sec:disc}

After completing the development, the authors analyzed the recurring empirical observations. A key conclusion is that keeping humans in the loop remains essential. However, under the shift-up approach, their focus can shift away from implementation details towards the following higher-level concerns.

    \textbf{Plan the big picture.} Performing true shift-up requires  plannig, defining and communicating the scope of the project. During code generation, supplying a high-level system overview, including requirements, high-level design, and sometimes even the folder structure and specific code, led to more coherent and relevant outputs. Updating and reusing this context periodically in prompts helped to maintain consistency. This reduces unwanted features and lowers the number of prompts needed for re-guiding the model, leading to improved workflow and smaller costs in prompting.

    \textbf{Enable prompt engineering.} LLMs generate outputs based on prompts, but crafting effective prompts is often challenging. Small differences in wording can lead to very different results, and instructions obvious to humans may be interpreted overly literally by the models. Since better prompts yield better results, experimenting with multiple variations is essential to refine the workflow. For true shift-up, prompt engineering becomes a critical skill, serving as the interface between human vision and GenAI capabilities.

    \textbf{Advance in small steps.} Advancing in small steps leads to clearer and more detailed outputs, as developers can implement each suggestion as it is generated and immediately detect issues in either the prompts or the results. However, in the tested models this approach proved insufficient for industry-scale shift-up. A related observation is that problems in prompts, workflows, and tooling must be addressed as soon as they arise, just as in traditional development. This need is reinforced by the tendency of LLMs to overlook failures in their own outputs, making continuous testing essential.

    \textbf{Manage versioning yourself.} Developers should be aware that LLMs may recommend outdated or deprecated tool versions. Since older versions can introduce compatibility issues and security risks, projects should always rely on up-to-date technologies. Consequently, developers’ expertise in version management and in assessing the compatibility of different tools remains essential. 

In summary, reaching shift-up needs a structured, incremental approach and an over-arching plan. 
The generated content also needs to be scrutinized by knowledgeable developers to find errors and deviations early and prompt the model to fix them.

\section{Future Plans}
\label{sec:conclusions}

The emerging GenAI native software  development has a significant impact on how software will be created in the future. In this paper, we propose that shift-up will fundamentally reshape the development altogether, implying a new GenAI native software development paradigm. In the extreme scenario,  developers do not write any of the code themselves, but all of the code in such a system would be under the responsibility of GenAI, and the developers need not understand its behavior, apart from requirements, quality of the implementation, and associated monitoring. 

Since using GenAI to assist software design and development is becoming prevalent among software developers, the software engineering community must start seriously studying and considering the consequences and challenges associated with the approach. In the advent of GenAI native software engineering, we should not consider GenAI as an omnipotent cult, but seek to better understand its role in coding, debugging, testing, integration and deployment. Some exact topics for research are listed below. Furthermore, because prompting is a key concept in GenAI assisted development, understanding where its limits lie is an essential step forward.

As demonstrated in the case study presented in this paper, GenAI is often treated as a separate activity rather than being integrated into the development flow. However, it is well understood that disruptions to the developers’ flow state can reduce productivity \cite{mikkonen2016flow}. Investigating how GenAI can provide assistance without breaking this state is therefore an important direction for future work. As a practical step forward, we plan to explore GenAI in controlled experiments, where new tools and methods that incorporate structured knowledge can be systematically benchmarked, helping to mitigate the biases that affect many of today’s studies.

In summary, it is still difficult to predict the long term impact of the disruption caused by
GenAI native software development. Transforming today's ad-hoc style use of GenAI into systematic practice is required to ensure the industry-grade maturity needed for wide adoption. This calls for new practices that result in the up-shift of the development focus.

\bibliographystyle{plain}
\bibliography{bibliography}

\end{document}